\documentclass{article}

\usepackage{arxiv}

\usepackage[utf8]{inputenc} 
\usepackage[T1]{fontenc}    
\usepackage{hyperref}       
\usepackage{url}            
\usepackage{booktabs}       
\usepackage{amsfonts}       
\usepackage{nicefrac}       
\usepackage{microtype}      
\usepackage{graphicx}
\usepackage{amsmath}
\title{Visual Abstraction}

\author{
  Ivan Viola \\
  King Abdullah University \\ of Science and Technology (KAUST)\\
  Saudi Arabia\\
  \texttt{Ivan.Viola@KAUST.edu.sa} \\
   \And
  Min Chen \\
  University of Oxford\\
  United Kingdom \\
  \texttt{Min.Chen@OERC.ox.ac.uk} \\
   \AND
  Tobias Isenberg \\
  Inria Saclay\\
  France\\
  \texttt{tobias.isenberg@inria.fr} \\
}

\newtheorem{defi}{Definition}
\newtheorem{theo}{Proposition}

\begin{document}
\maketitle

\begin{abstract}
In this article we revisit the concept of abstraction as it is used in visualization and put it on a solid formal footing. While the term \emph{abstraction} is utilized in many scientific disciplines, arts, as well as everyday life, visualization inherits the notion of data abstraction or class abstraction from computer science, topological abstraction from mathematics, and visual abstraction from arts. All these notions have a lot in common, yet there is a major discrepancy in the terminology and basic understanding about visual abstraction in the context of visualization. We thus root the notion of abstraction in the philosophy of science, clarify the basic terminology, and provide crisp definitions of visual abstraction as a process. Furthermore, we clarify how it relates to similar terms often used interchangeably in the field of visualization. Visual abstraction is characterized by a conceptual space where this process exists, by the purpose it should serve, and by the perceptual and cognitive qualities of the beholder. These characteristics can be used to control the process of visual abstraction to produce effective and informative visual representations.
\end{abstract}


\section{Definitions}
The term \textit{abstraction} often lacks a precise definition in many fields.
While several fields have defined the term for their own purposes, there is only a vague understanding of its meaning that is shared by all fields.
Some scientific disciplines and scholarly fields have adjusted the vaguely understood meaning to fit the needs of the respective discipline or field. 
In this article we first present our key definitions related to visual abstraction, and we then provide the justification for the definitions. 
In giving these definitions, we revise our previous set of definitions relating to the concept of abstraction~\cite{Viola2018}, based on new discussions related to, and insights from, our further literature study. Terminology related to abstraction has been adopted from  Lepp\"anen~\cite{Leppanen2007} and is discussed in \autoref{sec:ontology}.

\begin{defi}
\label{def:abstraction}
	An \textbf{abstraction} is a process that transforms a \emph{source thing} into a less concrete \emph{sign thing} of the \emph{source thing}. Abstraction uses a concept of \emph{point-of-view}, which determines which aspects of source thing should be preserved in its sign thing and which should be suppressed. 
\end{defi}
\begin{defi}
\label{def:representations}
	A \textbf{data representation} is a sign thing that stands in digital form for a \emph{referent thing} from reality or another \emph{sign thing}, using data structures or \emph{concept things}. Similarly, a \textbf{visual representation} is a  \emph{sign thing} that stands for a \emph{referent} from reality or another \emph{sign thing} so that it can be visually perceived and cognitively processed by a human observer.
\end{defi}

\begin{defi}
\label{def:visual_abstraction}
    Visual abstraction is a particular type  of abstraction where the sign thing is visual, while the source thing is either non-visual or visual. 
	A visual representation results from a process of \textbf{visual abstraction} if such transformation intentionally disregards certain aspects of data representations.
\end{defi}

\begin{defi}
\label{def:point-of-view}
The abstraction process also involves a \textbf{point-of-view} component defined through the task, which the visualization process aids to accomplish. This task is represented as a combination of \emph{targets} on which particular \emph{actions} are performed.  
\end{defi}

\begin{theo}
\label{def:quantifying_abstraction}
	The amount or \textbf{significance of abstraction} of a thing can be, in computer or signal representations, quantified by means of information theory.
\end{theo}
\begin{defi}
\label{def:meaningful_abstraction}
	A \textbf{meaningful visual abstraction} is a visual abstraction such that, for a given point-of-view and for a given \emph{purpose} or \emph{goal}, key aspects of the underlying \emph{referent thing} are preserved in the visual representation so that the cognitive load when perceiving it as a stimulus is significantly reduced. 
\end{defi}

\begin{defi}
\label{def:visualization}
	A \textbf{visualization} is a process that transforms data representations of a thing from reality into visual representations. Visualization is a process that is intended to be a meaningful visual abstraction process. The designers of visualization processes must understand the point-of-view component and tasks. Otherwise, they would not reach the full meaningfulness intended.
\end{defi}

\begin{defi}
\label{def:abstraction-axis-and-space}
	An \textbf{abstraction axis} is the \emph{perceived} sequence of visual representations that is assembled by the designer of a visualization system to illustrate a given point or series of points about reality. Each of the building blocks of an abstraction axis is the result of an individual abstraction process to a visual representation. Each transition between two successive abstraction axis building blocks can but does not have to remove information, some can also both remove and add information based on chosen blocks specific abstraction. If two or more abstraction axes are constructed such that they affect independent aspects of the visual representations, they can be combined into an \textbf{abstraction space} that observers can explore.
\end{defi}

\section{Flavors of Abstraction}
The notion of what is abstract and what is concrete is a fundamental discussion in philosophy, without a clear consensus. In its simplest terms, an abstract object has no physical referent, while concrete objects have physical referents. 
Reiterating Frege's writings, ``The Thought''~\cite{Frege1918} is even stronger in restricting what an abstract thought is: ``An object is abstract if and only if it is both non-physical and non-mental.'' An object is acknowledged as mental when ``it exists at a time if and only if it is the object or content of some mental state or process at that time.'' This statement implies that an abstract object is an object if and only if it cannot be found in nature, cannot be constructed, and one cannot even form a mental image of it.

Another definition of abstract objects is that they lack causal powers~\cite{StanfordEOP}. This means that abstract objects cannot affect other objects in any way. An empty set is such a case of an abstract entity as it does not have any causal powers.
The definition of abstract entity is often so strict that some philosophers deny the existence of an abstract entity as such. 
However, there seems to be better agreement on what an abstraction is: ``It is a distinctive mental process in which new ideas or conceptions are formed by considering several objects or ideas and omitting the features that distinguish them''~\cite{StanfordEOP}.
Lewis~\cite{Lewis1986} proposed that ``abstract entities are abstractions from concrete entities. They result from somehow subtracting specificity, so that an incomplete description of the original concrete entity would be a complete description of the abstraction.''
In the rest of the article we use the term abstraction aligned with these definitions to only describe a process, as we have also done in our own definitions at the beginning. The entity after abstraction is, in our case, denoted as a representation. We do not enter the dispute of whether it is an abstract entity or not. In such a way we build on the part that philosophers agreed upon, while we avoid the terminological controversy. Before we look at the use of abstraction in visualization, let us first consider its occurrence in related arts and sciences.


\subsection{Abstraction in the arts}
In the arts, the term \emph{abstract art} refers to non-figurative artwork, where the intent is to develop art beyond depiction of natural or man-made objects. The composition may exist with a degree of independence from visual references in the world~\cite{Arnheim1969}.
This art movement started during early 20\textsuperscript{th} century and emerged from figurative art. Artists such as Picasso, Mondrian, Kandinski, and many others originally depicted natural objects.
The beginning of non-figurative art started with a deep analysis and observation of the creative process, where the graphical elements that composed the rendering became themselves the subject of study.
The natural objects were gradually represented through collection of simpler geometric primitives. The artists searched for an expression of minimal set of visual elements that is still able to carry the figurative meaning. But they did not stop there. Artists further experimented with the graphical elements beyond recognizability of any corresponding figure from the rendering itself. Interestingly, one can sometimes discover a correspondence to their earlier works where a particular figure is still recognizable, thus transitively the figure can be imagined in the fully abstract art with such aids as well. It indicates that the artists still had a particular figure in mind, when rendering a particular art, while, without the prior work context, this figure would not be discovered by another human observer.
This gradual process, which transformed figurative art into what is now called abstract art, is abstraction.

\subsection{Abstraction and generalization in cartography}
In cartography, depending on a chosen scale for a map and its type/target audience, a subset of information is selected, the elements to be depicted are simplified and their depiction is adjusted. For example, streets can be shown with a much larger width than in reality, yet fine details of their path are removed. When zooming out, important elements and landmarks in the map are depicted, while generally less relevant elements are suppressed. At a particular level of scale, for example, the post office, a religious place, a building of historical significance, a bridge over the river, or the main streets are clearly depicted in the map, while similar objects in terms of spatial dimensions are abstracted into very simplified representations, if they are shown at all. The field has created a solid vocabulary and guidelines on how certain elements should be depicted and when should they be visible. In cartographic visual language, the umbrella term for guidelines of how different scales should depict certain information is~\emph{map generalization}~\cite{Buttenfield:1991:MGM}. We discuss the specific meaning of the term \emph{generalization} below, but other principles such as \emph{grouping} or \emph{classification} are applied here as well. In prior work, these concepts are considered as distinct abstraction principles and we discuss their specifics below.

\subsection{Abstraction in shape analysis}
In shape analysis, the term \emph{abstraction} typically refers to a skeletonization or extraction of topological features that represent  essential characteristics of the underlying shape~\cite{Cornea2005,Isenberg:2004:CTE}. Here, abstraction preserves the key properties of the geometric components such as their connectivity. The levels of detail of these abstracted representations are controlled through measures like persistence: this measure determines which structures are too small for particular scale to justify their validity and which are grouped into other larger-scale structures. Such abstracted representations facilitate the extraction of hierarchies in shapes to facilitate geometric linkage, multi-scale representations, and---importantly---the topological representation is much sparser and facilitates an unobstructed clear view on the key geometric properties. The same holds for the topology of flow data, where a flow field is \emph{classified} into points and regions of certain uniform properties such as sinks, sources, and separatrices (curves or surfaces) that partition the flow according to its long-term behavior.

\subsection{Mathematical abstraction}
The term mathematical abstraction refers to a process of transforming a specific real-world situation into generalized form using mathematical formalism.\footnote{https://en.wikipedia.org/wiki/Abstraction\_(mathematics)} The specifics which do not affect the solution to a given problem are removed so that, in the end, only a set of key elements with properties and relations to each other remains, which can be expressed formally. Problems to solve in mathematics class are frequently expressed as real-world situations. The tasks are to abstract from the real-world specifics and apply a mathematical formalism that provides the answer to the given problem. The development of mathematics and physical sciences has advanced through mathematical abstraction into Euclidean geometry, algebra, and analysis. These developments have been possible due to humans being capable of thinking in an abstract way.

\subsection{Abstract thinking}
School students are trained in abstract thinking by being challenged to solve a specific real-world problem. To be able to do so, they are trained to abstract from the case specifics by extracting only the essential components so that a formal solution can be calculated and, finally, interpreted back for the specific real-world scenario. Abstract thinking is, according to cognitive psychology~\cite{Mosby2012}, the most complex stage in the development of cognitive thinking, where generalizations and concepts are used in the thought process. From a set of observations, hypotheses can be formed and logical reasoning can lead to conclusive statements~\cite{Mosby2012}.

\subsection{Abstraction in object-oriented design}
In computer science, the term abstraction achieves yet another flavor of its meaning. In object-oriented design, the most frequently used programming methodology, it primarily relates to the definition of classes and methods that cannot be instantiated. 
Typically, classes and methods are hierarchically grouped into increasingly abstract constructs such that implementations of particular functionality can be shared among many different elements.
While for most of these classes it is possible to create instances, an abstract class is a construct that itself cannot be instantiated but which organizes the functionality into a comprehensive representation. The class hierarchy as the outcome of such abstraction gives a clear understanding of differences in functionality among various classes as well as what they have in common. It also facilitates further extensibility of existing code to support new cases that were not considered in the initial software design.

\subsection{Abstraction ontology}
\label{sec:ontology}
In the area of information and knowledge modeling, a particularly interesting past work closely relates to our own investigation. Lepp\"anen~\cite{Leppanen2007} distinguishes between first-order and second-order abstraction. First-order abstraction is associated with primary things, while second-order abstraction acts upon a predicate that defines the primary things. An example of a primary thing is \emph{sedan} with several predicates, among others a \emph{color}. The result of the abstraction of a sedan would be a \emph{car} or a \emph{vehicle}, which corresponds to first-order abstraction. Let us assume that an instance of a sedan is painted with a particular blue, for example \emph{Maya Blue}. This predicate can also be abstracted to \emph{light blue} or \emph{blue}, a process which is of the second-order abstraction type and is also termed as predicate abstraction.

Importantly, Lepp\"anen defines four elementary abstraction principles: classification, generalization, composition, and grouping. First, classification is defined through the term \emph{isInstanceOf} or that instances are \emph{typeOf}. The opposite to classification is instantiation. Second, generalization is a principle of abstraction where the differences of subtypes are suppressed to fit a supertype. This refers to an \emph{isA} relationship and the antonym to generalization is specialization. Third, composition is a principle of abstraction in which a whole concept is composed of part concepts. These parts are abstracted to form a whole object. This refers to a \emph{partOf} relationship and its opposite is the decomposition. Finally, the last principle of abstraction is grouping which relies on a \emph{isMemberOf} relationship and whose opposite is individualization. For example, a particular person can be a member of a political party. This abstraction includes aggregation, set membership, and association. Both, first-order and second-order abstractions can benefit all four elementary abstraction principles. In all cases, an important property to highlight is that abstraction is associated with an intentional and controlled loss of information.

Lepp\"anen's work stresses the importance of the concept \emph{point-of-view} that plays crucial role during the abstraction process. When using classification on a \emph{thing} termed, for example, \emph{Margaret Thatcher}, the abstraction along classification would lead to entity \emph{female} or \emph{UK Prime Minister}. If we would be using grouping, the abstraction would lead to \emph{Conservative Party}. In case the composition principle is used for abstraction of \emph{UK Prime Minister}, the outcome would be \emph{UK Government}. Therefore, things might generally have many different kinds of abstractions as things from reality are typically embedded in a complex and intertwined abstraction hierarchies.

In his work, Lepp\"anen combines philosophical and semiotic standpoints. In the context of semiotic frameworks~\cite{Ogden1923}, they refer to three kinds of \emph{things}: a \emph{concept thing}, a \emph{referent thing}, and a \emph{sign thing}. Concepts are mental constructs, words of mind, and form basic components of human knowledge. A referent is an element of reality that relates to the concept. A reality describes a set of anything that exists or can possibly exist, physically or virtually. A sign is anything that can stand for something else, including symbols, text, or images. As such it is a representation of a concept. These concepts are used below in the discussion of abstraction in visualization. We applied the same terminology in our definitions from the start of the article, but we added the concept of a \emph{source thing} (\autoref{def:abstraction}).

\subsection{Summary of abstraction in the world outside visualization}
The intuitive understanding of abstraction has been reinforced by this brief excursion into various fields and that stand and argue for abstraction. We can observe that the term is not used uniformly and that it is frequently exchanged with other terms. The recurrent pattern is that abstraction relates to formation of some higher-order constructs or representations that are result of a transformation of lower-level entities. The lowest entities are more tangible, while the higher levels of the abstraction hierarchy are further removed from tangibility and become more mental constructs and concepts (defined as \emph{the constituents of thoughts}~\cite{StanfordEOP}) that, in one way or another, allow humans to recognize certain characteristics clearer than the lower-level representations. The ability to abstract seems to be one of the core properties of humans, present while shaping the entire body of analytical knowledge humankind has formed throughout our history.

\section{Abstraction for Visualization}

Let us now investigate how abstraction manifests itself in visualization. We propose that abstraction is equally central to visualization as it is to other areas in which analytic reasoning is the core part of a processing workflow.
Visualization is the process of transforming the digital representation of data into visual representations that are exposed to a human viewer (\autoref{def:visualization}). It takes advantage of the fact that most humans are extremely efficient in comprehending information presented as a visual stimulus. Naturally, this stimulus has to be well designed to convey the intended information (\autoref{def:meaningful_abstraction}). This aspect is the main concern of the visualization mapping stage of the visualization pipeline.
Visualization is omnipresent in studying various real-world phenomena, conveying structures, methods, or concepts.
In visualization, the abstraction process guides the transformation into visual representations (\autoref{def:representations},~\ref{def:visual_abstraction}), similar to the process of abstract thinking. In some sense it serves as an extension of the working memory, where needed information can be instantaneously accessed.
We thus first clarify the meaning of abstraction in visualization and then discuss its core properties.

To bring visualization into the context of semiotic frameworks, the \emph{sign} is termed as \emph{representation}, both digital and visual, and the \emph{referent} is the studied phenomenon from \emph{reality} (\autoref{def:abstraction},~\ref{def:representations}). The \emph{concept} is what relates to the \emph{referent} and can be conveyed through the representation. In visualization, abstraction is performed at least in three stages: first, the abstraction of the reality into data representations and, second, the data representation is, through abstraction, transformed into visual representations. Third, a visual representation is transformed to a mental model or a memory representation through the perceptual and cognitive processes of the human observer.

Abstraction has occurred if the quantum of information before the abstraction is higher than in the representation after abstraction, while some aspects of the original representation are preserved and become more prominent (\autoref{def:abstraction}). In case there is no intended information loss, we refer to a more general term \emph{transformation} or \emph{mapping}. For example, several simultaneous abstraction processes that individually work on different aspects of the \emph{things} could be combined, some work in a positive direction (removal of information) and others in a negative direction. This could lead to composite \emph{transformation} or \emph{mapping} that transfer one representation into another, with information loss and information gain at the same time.

\subsection{Task abstraction} 
Visualization is driven by a particular \emph{intent}. There is a reason behind a visualization, even in the casual scenarios. This intent defines the \emph{point-of-view} (\autoref{def:point-of-view}), which, as a controlling mechanism, can steer how abstraction changes the representations. In the visualization literature, Munzner~\cite{Munzner:2014:VAD} describes a hierarchical framework into which specific individual visualization usage scenarios can be abstracted. On the highest level, Munzner classifies the tasks as a combination of an \emph{action} upon a \emph{target}. The \emph{action} class is instantiated into \emph{analyze}, \emph{search}, and \emph{query}, which can be further instantiated into lower-level classes of \emph{actions}. The \emph{target} is instantiated into \emph{data} in general, \emph{attributes}, \emph{networks}, and \emph{spatial data} which are further instantiated into more detailed targets. It is the combination of the \emph{action} and \emph{target} that would define the \emph{point of view} to guide the abstraction process.

\subsection{Data abstraction}
Munzner~\cite{Munzner:2014:VAD} also defines various types of \emph{data} and \emph{data sets} for visualization. Data types are \emph{items}, \emph{attributes}, \emph{links}, \emph{positions}, and \emph{grids}. Data set types are \emph{tables}, \emph{networks}, \emph{trees}, \emph{fields}, \emph{geometry}, \emph{clusters}, \emph{sets}, and \emph{lists}. All these types are \emph{concept things} (\autoref{def:representations}). The data abstraction here refers to the transformation from the real-world phenomenon, the \emph{referent thing}, into data structures (\emph{concept thing}) and digital representations (\emph{sign thing}), to facilitate an efficient and automatized computational processing. This task of data abstraction is somewhat similar to the mathematical abstraction process. In both cases we end up with a formal representation on which standardized mathematical or computational machinery can be applied.

The initial data abstraction is typically performed during the acquisition process. Either real-world observations are made and digitally stored in a particular data representation or even a mathematical model is formulated based upon these observations. Both forms are data representations abstracted from the thing that exists in reality, and these representations have been achieved through a classification process.

The result of the initial data abstraction is frequently further abstracted into another data representation to promote a particular \emph{point-of-view}, neglecting unimportant aspects of the original data representation. As such, the filtering operation is typically applied, which might be considered to relate to map generalization and as such corresponds to the generalization abstraction principle. Once the data representation contains the relevant data prominently, a conversion into data representation is performed that can efficiently be visually represented.

\section{Visual Abstraction}

\begin{figure}[t]
	\centering
	\includegraphics[width=0.6\textwidth]{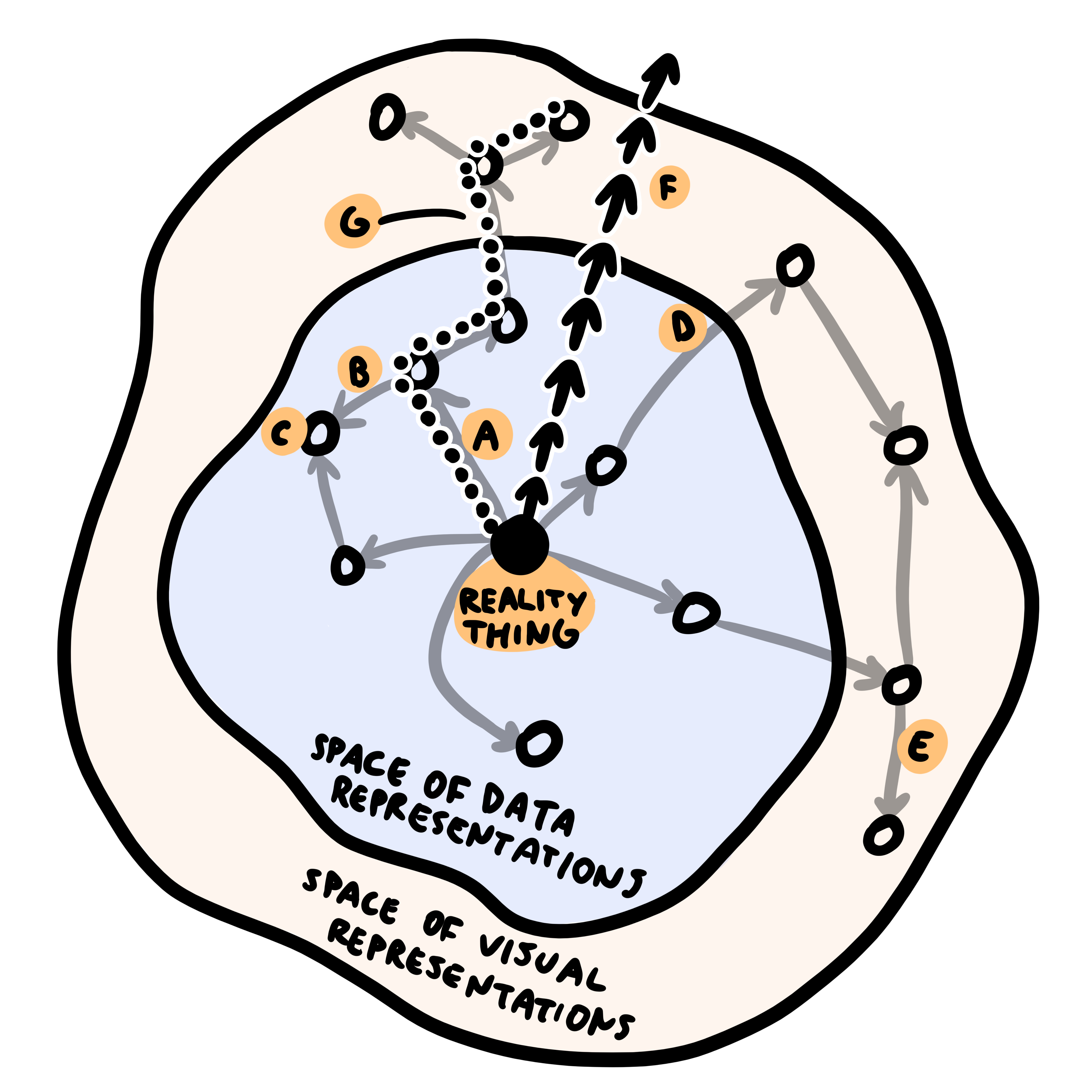}
	\caption{Abstraction space in which a \emph{thing} from reality is gradually transformed into visual representations: a) initial abstraction into a digital form, b) data abstraction into new data representation, c) different data abstractions can lead to identical data representation, d) visual abstraction transforms the data representation into a visual representation, e) visual abstraction transforms one visual representation into another visual representation, f) the abstraction space encodes less and less information from the original \emph{thing} from reality. The further from center the more sparse the representation is. g) the dotted line conveys a visualization pipeline that can be seen as a composite visual abstraction.}
	\label{fig:space}
\end{figure}

After series of data abstractions and transformations (the latter when no information loss happens), in visualization, the data is transformed\discretionary{/}{}{/}abstracted into visual representations. A visual representation is then shown on a display, perceived, and further cognitively processed by a human observer. The visual abstraction process that generates this visual representation can be performed in many ways: In the case of kernel density estimation plots or clustering techniques, for instance, data can be visually abstracted using a composition principle such that smaller elements become a part of higher-order representations. In case of volumetric scalar fields, the voxel values can be classified into color and opacity ranges. By this, some voxels become abstracted into types such as air, soft tissue, or hard tissue. Level of detail techniques would typically relate to composition or grouping; in atomistic visualization, individual atoms become member of particular molecules, which in turn become members of certain compartments, up to cells. In many cases of particular visual abstraction it can be simultaneously argued for different abstraction principles, and there might be more principles than those proposed by Lepp\"anen~\cite{Leppanen2007}.

Munzner~\cite{Munzner:2014:VAD} provides a conceptual framework according to which visual representations or encoding can be categorized. This framework is rather extensive, however on the low level the visual encoding can abstract data representations through two key aspects. The first aspect is the graphical \emph{mark} that positions each data element: \emph{points}, \emph{lines}, and \emph{areas}. These marks further encode quantitative information or their mutual relationship is conveyed through various perceptual channels: \emph{position}, \emph{shape}, \emph{color}, \emph{size}, and \emph{angle}. These basic low-level perception-driven visual elements can be combined to create rich spectrum of possible visual representations. These visual representations can be used to \emph{encode}, \emph{manipulate}, \emph{compare}, or \emph{reduce} the data in the visual representation space.

Data representations and visual representations can be ordered according to how much they abstract a particular phenomenon. The abstraction process is depicted in \autoref{fig:space}. We can see the abstraction space related to one \emph{thing}, one entity from reality. There are several ways how the \emph{thing} can be abstracted into a digital form. After this first stage of the process, the data representation has been abstracted from the \emph{thing}. There could be several data abstractions applied, under which the data becomes sparser and sparser so that the information sought by the user or intended by the visualization creator becomes gradually clearer. Sometimes even the series of abstractions can take different paths yet still result into the same data representation. In practice, however, such data abstractions would only apply to a given path to a particular visual representation, as most visualization systems will maintain their original datasets to allow users to also observe different visual representations, which would be the result of a different sequence of data abstractions.

After the sequence of data abstractions, the data representation is still non-visual. If we apply a visual mapping to such data (whether with intentional loss of information or not), we achieve a visual representation that can be viewed on a display. But even visual representations can be further transformed into sparser visual representations by means of visual abstraction. The more far away in the abstraction space, the less information from the reality is preserved. If we concatenate a path from the reality to the final visual representation, we can see a visualization pipeline. In case we perform a transformation so that the distance between the original representation and the reality and the target representation and the reality are the same, we do not perform an abstraction. If the target representation is closer to the reality than the original representation, we perform an inverse operation to abstraction. Yet this inverse abstraction only happens in the eyes of the beholder, as we always remove information along the path from reality to visual representation.

\subsection{Meaningful Abstraction}
It is
not clear whether an abstraction has to be meaningful or whether its only condition is a loss of information. What if, for example, a high-dimensional data set is projected onto fewer dimensions? Projection is, in principle, a valid abstraction. But what if we project only every second data element and create a confusing data representation in which only half of the data set is projected onto lower-dimensional space. Is such a meaningless projection also an abstraction? From the information-theoretic point of view we have lost a certain amount of information, so it can be considered as an abstraction. To differentiate us from this view, we should define the term \emph{meaningful abstraction} for those abstractions that are useful in some application contexts (\autoref{def:meaningful_abstraction}).

Visual mapping may result into a representation with an equal amount of information, however, more visually confusing than the previous representation. For example, it is known that humans have difficulties with identifying portraits of known faces if they are rotated by 180 degrees from the natural portrait orientation~\cite{Thomson:1980:MTN}. From the information-theoretic point of view, the rotation does not remove information from the image, but there is a significant difference in cognitive load between these two representations. Such a rotation is consequently not a meaningful visual mapping. The same holds for two visual representations of a graph, a node-link diagram and an adjacency matrix. When the one visual representation is transformed into another, no information is lost. Yet the cognitive load for viewers differs between these two representations. Building on the term of \emph{meaningfulness}, a visual representation can be more meaningful (or effective) for a particular intent than another visual representation. Visual abstractions that lead to these representations might be ordered or perhaps even quantified in how meaningful they are. 

The concept of \emph{meaningfulness} in terms of visual abstraction processes is tightly related to visual perception processes. In principle, a meaningful visual abstraction makes the job of visual processing simpler so that less of a cognitive processing needs to be invested, for a given purpose or goal, in comprehending the abstracted visual representation to understand the intended aspects of the reality. Therefore, visual abstractions relevant for visualization will need to result into lower cognitive load when comprehending the abstracted representation, for the chosen intention. Therefore the \emph{meaningful visual abstraction} has to pass two conditions: the target visual representation has to formally contain less information and the cognitive load has to be lower. The perceived information, if not increase, should decrease at most linearly with the cognitive load.

At this point we solidify the previous discussion and define some key terms in visualization.
Abstraction is a process, it is a transformation along which some information is intentionally lost to give prominence to the \textbf{higher-level} information within. The abstraction process results into a representation. For pure data abstraction, it results in a data representation, while, when visual abstraction is involved, it results in a visual representation. These abstractions can be considered as meaningful as long as they are benefiting particular application example or purpose. The meaningfulness property is scoped by the set of meaningful applications.
A visual representation is the result of a visual transformation. When information is intentionally lost and the cognitive load is lower, while the perceived information loss is, at most, linear with the cognitive load difference, then we consider the visual abstraction as meaningful. 
Visual mapping and visual encoding, while both having their distinct meaning, can be used interchangeably with visual transformation. Visual metaphor operates on the concept of \emph{analogy}. It presents a sign thing of a different referent thing from reality than the one originally regarded. This way visual mapping associates properties of one referent thing to another referent thing. An example of a visual metaphor are Chernoff faces, where different facial properties encode multivariate data~\cite{Chernoff1973}.

\subsection{Abstraction Axes and Abstraction Spaces}

So far we mainly discussed the process of abstraction from reality via data representations to visual representations. Yet we also showed that positive or negative abstraction can be \emph{perceived} by a viewer as he or she is manipulating this abstraction chain or visualization pipeline. For better describing the latter aspect, Viola and Isenberg~\cite{Viola2018}, inspired by earlier examples in visualization~\cite{Miao2018b,Miao2018a,Mohammed2018,Zwan2011} as well as in the arts world, proposed the notion of \emph{axes of abstraction} which could form an \emph{abstraction space}. With these two concepts we can describe the abstraction that is perceived and controlled by the beholder, in contrast to the abstraction that is applied as a particular visual representation is generated (\autoref{def:abstraction-axis-and-space}).

An \emph{abstraction axis} in this concept is the previously mentioned virtual, perceived connection between different end points of the previously discussed abstraction process. This connection arises for observers as they adjust the settings of the visualization pipeline. This notion, however, assumes that, for each abstraction axis, there is a clearly identifiable succession of changes to the visual representation that (a) decreases the amount of information in each step and (b) provides a meaningful generalization of the depicted content to the viewer. In fact, Viola and Isenberg~\cite{Viola2018} even state that abstraction axes do not need to be unique: for their chosen example from structural biology~\cite{Zwan2011} they show that a molecular van-der-Waals surface-based molecular representation can be subjected to two alternative forms of structural abstraction (a phenomenon they call ``forking'' of axes)---one leading to a surface-based abstraction via different probe sizes and one leading to a second-order representation via balls-and-sticks, licorice (sticks-only), and backbone representations.

In particular for this latter form of abstraction, one could argue that condition (a) is not necessarily met: while the transition from van-der-Waals surfaces to the licorice representation in van der Zwan's~\cite{Zwan2011} model certainly removes the detail of the graphical atom representations, it simultaneously also adds representations of the bonds between atoms that did not exist in the starting configuration: the representations of atoms with implicitly represented bonds are continuously replaced with representations of bonds with implicitly represented atoms. One could thus argue that in this transition no abstraction happens, only one representation is smoothly transitioned into another. This transition, however, only happens in the eyes of the beholder; at any given point along the transition still abstraction happens from reality (source things) via data representations to visual representations.

A more recent example is the work by Miao et al.~\cite{Miao2018a} who similarly constructed a progression of transitions from an atom-based representation of DNA nanostructures and the mechanical building blocks of the nanostructures to be built. Interesting in their progression of ten abstraction stages is that, while the first and last are fairly clear, the specific order of the sequence in-between is not and was created based on the discussions with and needs of their collaborating domain scientists.

Based on these two examples, we thus suggest that pure and continuous abstraction axes are rather rare. Instead, abstraction axes are typically composed of smaller building blocks where one representation is (typically) seamlessly transformed into another and as such forms a constructed sequence. In practice we often find abstraction axes that progress from a representation with more information to a representation with less detail, thus the name abstraction axis. We can also find transitions, however, that remove one type of visual detail and replace it with another type of visual information. Abstraction axes are always constructed with a given purpose and application case in mind and are not unique. If two abstraction axes work on independent attributes of the visual representation and can thus be independently controlled, then they form an \emph{abstraction space} (\autoref{def:abstraction-axis-and-space}).


\section{An Information-Theoretic Analysis of Abstraction as a Process}
\label{sec:InfoT}
As described above, the notion of ``abstraction'' encompasses a wide range of definitions in different contexts.
It can be quite difficult for a single mathematical formulation to encapsulate the essences of these definitions.
In this section, we examine the characteristics of the process of abstraction using information-theoretic measures.
We show that the definitions given at the beginning of this article can be explained using an information-theoretic metric, which therefore offers a potential means for modelling and measuring visual abstraction.

\begin{figure}[t!]
\centering
\begin{tabular}{@{}c@{\hspace{2mm}}c@{}}
\includegraphics[height=36mm]{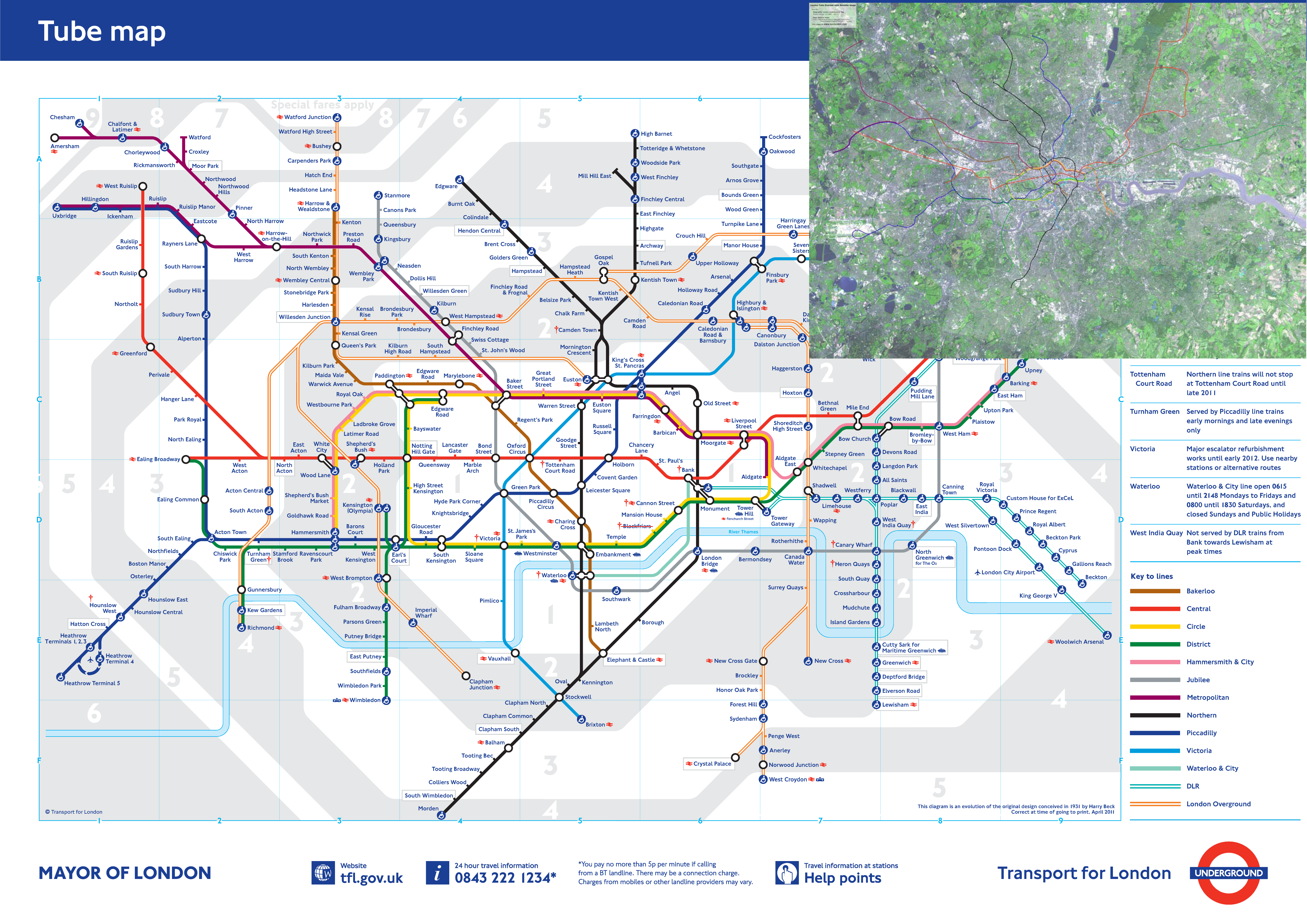}
& \includegraphics[height=36mm]{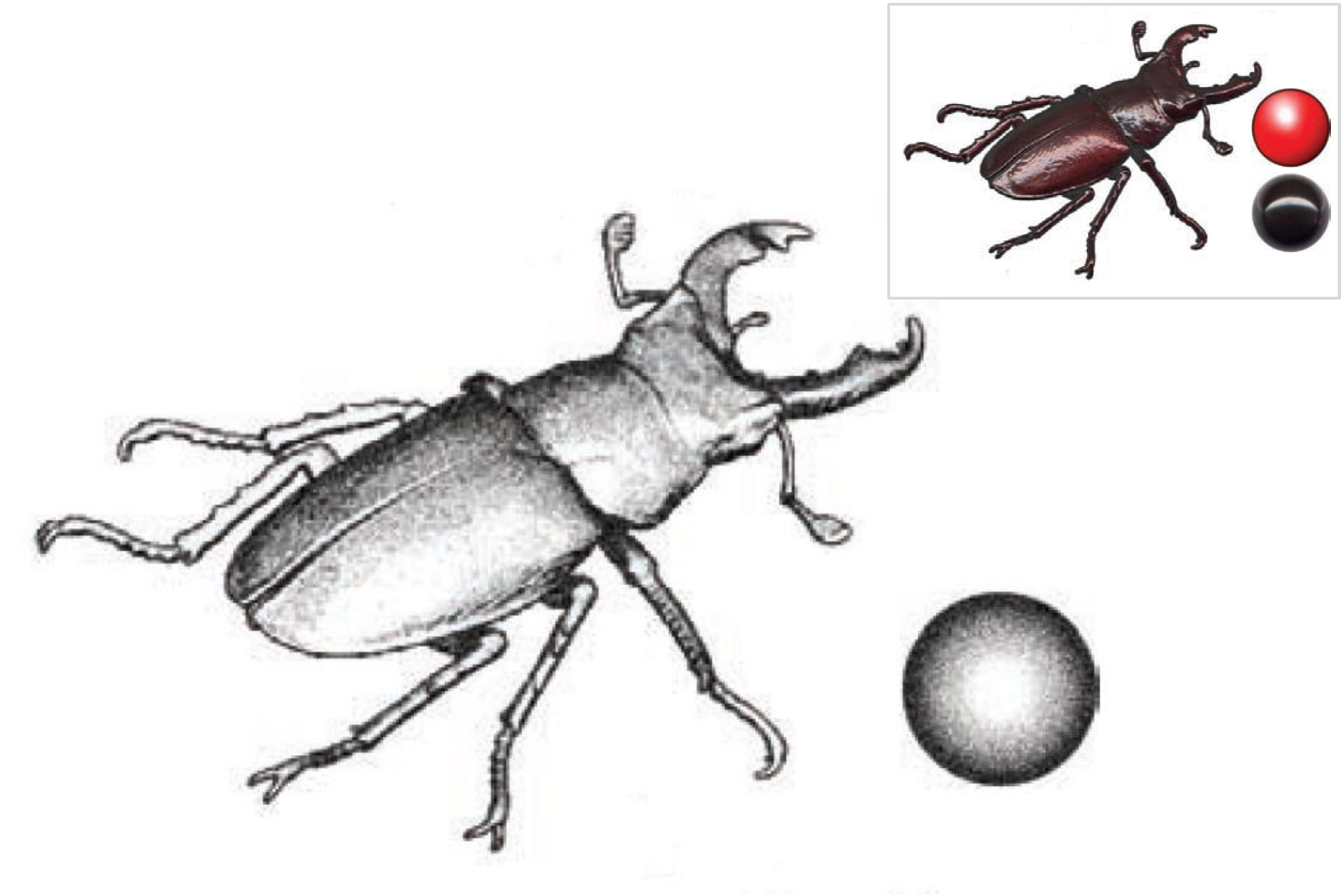}\\
(a) London underground map~\cite{TfL:2018:map} &
(b) pen-and-ink volume rendering~\cite{Bruckner2007}\\[2mm]
\includegraphics[height=36mm]{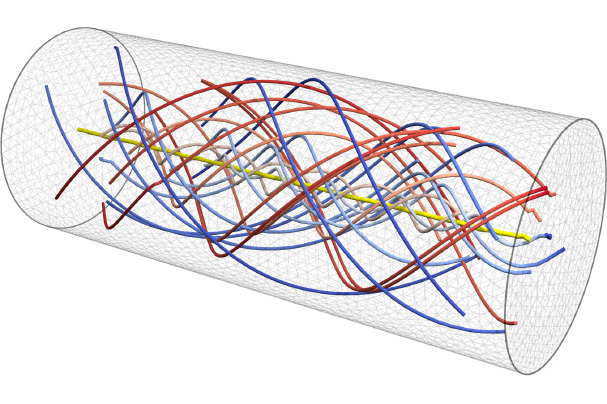}
& \includegraphics[height=36mm]{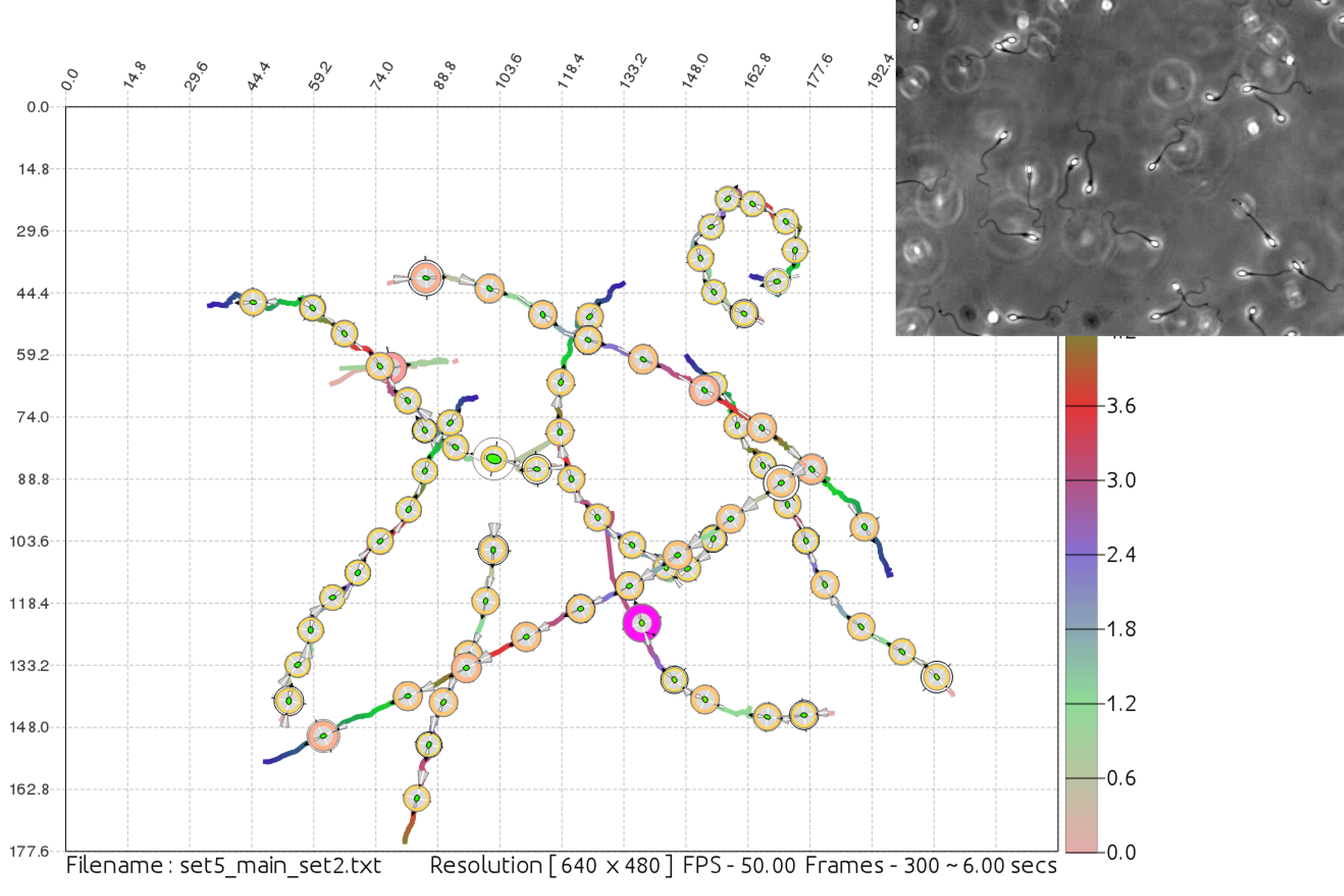}\\
(c) 3D streamline flow visualization~\cite{Oster:2018:CGF} &
(d) glyph-based video visualization~\cite{Duffy:2015:TVCG}\\[2mm]
\includegraphics[height=36mm]{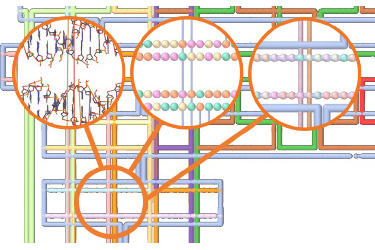}
& \includegraphics[height=36mm]{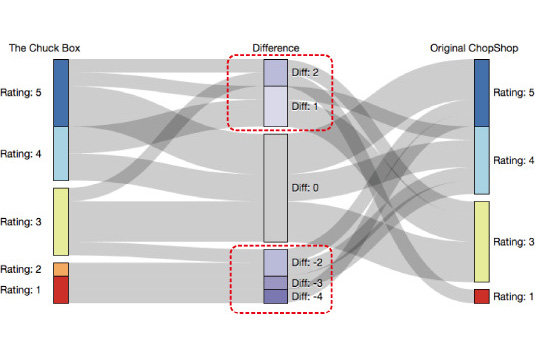}\\
(e) network visualization~\cite{Miao2018b} &
(f) parallel coordinates visualization~\cite{Wang:2018:CGF}\\[2mm]
\includegraphics[height=36mm]{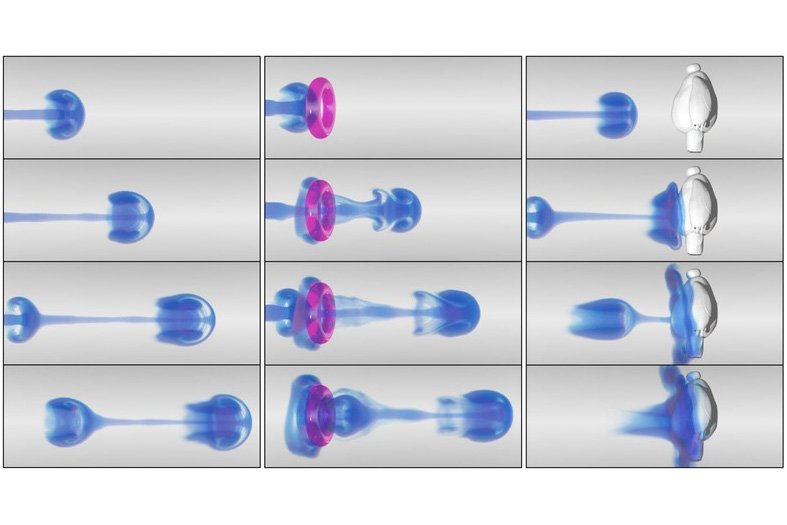}
& \includegraphics[height=36mm]{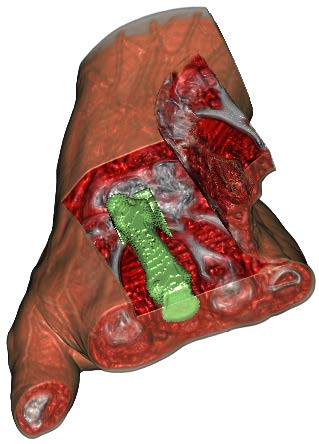}\\
(g) 3D flow visualization~\cite{Zhai:2017:CGF} &
(h) volume visualization and deformation~\cite{Birkeland2009}
\end{tabular}
\caption{Examples of visualization images that may attract different views as to whether they are the results of visual abstraction processes. Most would agree that (a)-(d) are visually abstract, many would content that (e) and (f) are considered so, and some might be hesitant about (g) and (h).}
\label{fig:ABS-examples}
\vspace{-2mm}
\end{figure}

\autoref{fig:ABS-examples} shows several visualization images generated using some typical visualization techniques.
Most visualization researchers would unreservedly refer to the first four images, (a)--(d), as results of visual abstraction, and many would contentedly accept a suggestion that (e) and (f) are also results of visual abstraction, but some would be hesitant to consider (g) and (h) as such abstracted representations.
Nevertheless, one can also argue that the latter four images, (e)--(h), are also results of visual abstraction because, in comparison with the source data, some information has been abstracted away and, in comparison with statistical abstraction of the source data, the information presented is visual.

First, the level of willingness for people to consider a visualization image as an abstracted visual representation does not appear to be related to the quality of the image or the usefulness of the technique that generates the image.
Second, we can observe that both \autoref{fig:ABS-examples}(a) and \autoref{fig:ABS-examples}(h) feature some deformation, and deformation does not seem to be a critical factor that influences the perception of visual abstraction results.
Similarly, from a comparison of (b) vs. (h) and (c) vs. (g) we can observe that the types of data to be visualized do not have decisive influence upon the perception of the term ``visual abstraction''.
Third, we can also observe that an impression of photorealism or just a perceived intention seems to bring about the hesitation in characterizing a visualization image as the result of visual abstraction.
Meanwhile, having no or less photo-realistic effect in an image (e.g., (e) or (f)) does not immediately imply visual abstraction either, at least to some people. Here the adjective ``photo-realistic'' indicates that the rendering algorithm used was designed to achieve a photo-realistic effect, without implying that the image resulting from the rendering process actually resembles a photograph.

One hypothesis is that our willingness or hesitation to consider a visualization image as resulting from visual abstraction relates to an unconsciously-integrated reasoning about two conditions of visual abstraction.
\renewcommand{\theenumi}{\Alph{enumi}}
\begin{enumerate}
\item \textbf{A visual abstraction is a transformation from data to its visual representation with some information loss}---Here data can be of any data types including visual data (e.g., image corpora and videos).
This can be considered as a broad definition of visual abstraction, and encapsulates the aforementioned definitions in cartography and shape analysis.
While introducing a constraint of visual output, it exhibits a parallel with the definitions in relation to mathematical abstraction, abstract thinking, and grouping in object-oriented design.
All eight images in \autoref{fig:ABS-examples} satisfy this condition in general. We will discuss information loss in detail later.
%
%
\item \textbf{A visual abstraction is a transformation from a more photo-realistic visual representation to a less photo-realistic one}---This can be considered as a narrow definition of visual abstraction, and encapsulates the aforementioned definitions in art, cartography, and shape analysis.
It applies to transformations with visual input as well as visual output. Considering the examples in \autoref{fig:ABS-examples}, for images (a), (b), (c), and (d), it is relatively easy for one to imagine their photo-realistic counterparts.
Although some of these images can be generated directly from source data that may not be visual, a subjective impression of a transformation that decreases photorealism is sufficient for viewers to associate these images with abstraction.
Meanwhile, it is harder to imagine a photo-realistic version of (e) or (f), and therefore this condition does not appear to be applicable to them. 
For images (g) and (h), it is intuitive to consider them more photo-realistic than less. They not only fail to satisfy, but also negate, this condition. 
\end{enumerate}

We can easily see that reading data using a spreadsheet or reading their statistical summary do not meet either condition.
Images (a), (b), (c), and (d) in \autoref{fig:ABS-examples} satisfy both conditions.
Images (e) and (f) satisfy condition A but not B.
Images (g) and (h) satisfy condition A but negate B.
Suppose that we had a numerical score 2 for condition A, score 1 for condition B, score 0 for not applicable, and score $-1$ for negation.
Spreadsheet or statistical summary would score 0; (g) and (h) would score 1; (e) and (f) would score 2; (a), (b), (c), and (d) would score 3.
Such a scoring system would reflect the level of willingness for one to characterize a visualization image as the result of visual abstraction. 

We can also infer that condition A is more essential than condition B.
Without A, images (e) and (f) would not be considered as results of visual abstraction at all.
Without B, there would not be any hesitation about whether images (g) and (h) are results of visual abstractions.

However, condition A does not in itself meet the expectation for the minimal quality that the process of meaningful visual abstraction should possess, since arbitrarily throwing away information should not be referred to as meaningful abstraction.
Below we use several information-theoretic measures to clarify Condition A.

Let $P_{d \rightarrow v}$ be a process for transforming a dataset $d$ to a visualization image $v$.
Let $\mathbb{D}$ be the data space containing all possible datasets that $P_{d \rightarrow v}$ can take as its input, and $\mathbb{V}$ be the data space containing all possible visualization images that $P_{d \rightarrow v}$ can generate.
In information theory,  $\mathbb{D}$ and $\mathbb{V}$ are referred to as \emph{alphabets}.
The dataset $d$ is thus a letter in the input alphabet $\mathbb{D}$, and the visualization image $v$ is a letter of the output alphabet $\mathbb{V}$.
The process $P_{d \rightarrow v}$ can thus be written as $P_{d \rightarrow v}: \mathbb{D} \longrightarrow \mathbb{V}$.

The Shannon entropy measures the amount of uncertainty or variation of an alphabet.
Let $p(d)$ be the probability of a dataset $d$ in the context of an application. The Shannon entropy of $\mathbb{D}$ is thus defined as:
\[
    \mathcal{H}(\mathbb{D}) = - \sum_{d \in \mathbb{D}} p(d) \log_2 p(d)
\]
When all letters in $\mathbb{D}$ have the same probability, we have
$\mathcal{H}(\mathbb{D}) = \log_2 \| \mathbb{D} \|$, where $ \| \mathbb{D} \|$ is the number of different letters in $\mathbb{D}$.
Similarly, we can measure the Shannon entropy $\mathcal{H}(\mathbb{V})$  as the amount of uncertainty or variation of $\mathbb{V}$.

\emph{Alphabet Compression}~\cite{Chen:2016:TVCG:78} is the difference $\mathcal{H}(\mathbb{D}) - \mathcal{H}(\mathbb{V})$, which is a coarse indication of the amount of information loss of the visualization process $P_{d \rightarrow v}$.
Consider a simple example. $\mathbb{D}$ is defined by a real variable, $X$, which may take valid values between 0.00 and 10,000.00 at two decimal point precision. There are thus 1,000,001 possible values. Let all values have the same probability. We have $\mathcal{H}(\mathbb{D}) \approx 20$ bits.

Meanwhile, we consider a process $P_{d \rightarrow v}$ that plots a value $d \in \mathbb{D}$ as a bar in a single-variable bar chart using a canvas with 1000H $\times$ 100W pixels. The maximum resolution available for the mapping function $P_{d \rightarrow v}: \mathbb{D} \longrightarrow \mathbb{V}$ is 1000 pixels, thus 1001 bar charts with different bar heights. We have $\mathcal{H}(\mathbb{V}) \approx 10$ bits.
The alphabet compression is therefore about 10 bits. In terms of Condition A, there is about 10 bits of information loss. Therefore, any visualization process, which features many-to-one mapping from data to visual objects, typically exhibits positive alphabet compression.
Only when the variation of $\mathbb{D}$ is very small, e.g., using the above canvas to plot an integer variable in the range of [0, 100], the amount of alphabet compression can be zero. In the worst scenario, the plotting function randomly depicts a bar with a height between 0 and 1000 pixels, the amount of alphabet compression would be negative.

All images in \autoref{fig:ABS-examples} feature many-to-one mappings.
For example, the distortion in \autoref{fig:ABS-examples}(a) is a kind of many-to-one mapping, since many potential track layouts would lead to the same metro map.
In the image rendered with a pen-and-ink effect in \autoref{fig:ABS-examples}(b), each white pixel could be a placeholder for many differently colored pixels that have been abstracted away.
Each glyph in \autoref{fig:ABS-examples}(d) is a very low resolution visual representation of some 20 values, most of which are real numbers.
In the volume-rendered image in \autoref{fig:ABS-examples}(h), each pixel results from a rendering integral that transforms a few hundred voxel values to an RGB trio. Many different combinations of these voxel values could result in pixels with the same color.

Hence, a process for generating visualization images from relatively complex datasets features many-to-one mappings, which means information loss or positive alphabet compression.
According Condition A, such a process is thus a process of visual abstraction.

However, what quantifies a visualization or a meaningful visual abstraction must be a process that is intended to generate ``meaningful'' visualization images from input datasets. The word ``meaningful'' implies three factors: (i) the viewer can interpret what is being depicted; (ii) the viewer's interpretation of what is depicted is reasonably correct in relation to the original data; and
(iii) the viewer's interpretation errors due to information loss do not have serious impact on the viewer's task.

\begin{figure}[t]
\centering
\includegraphics[width=100mm]{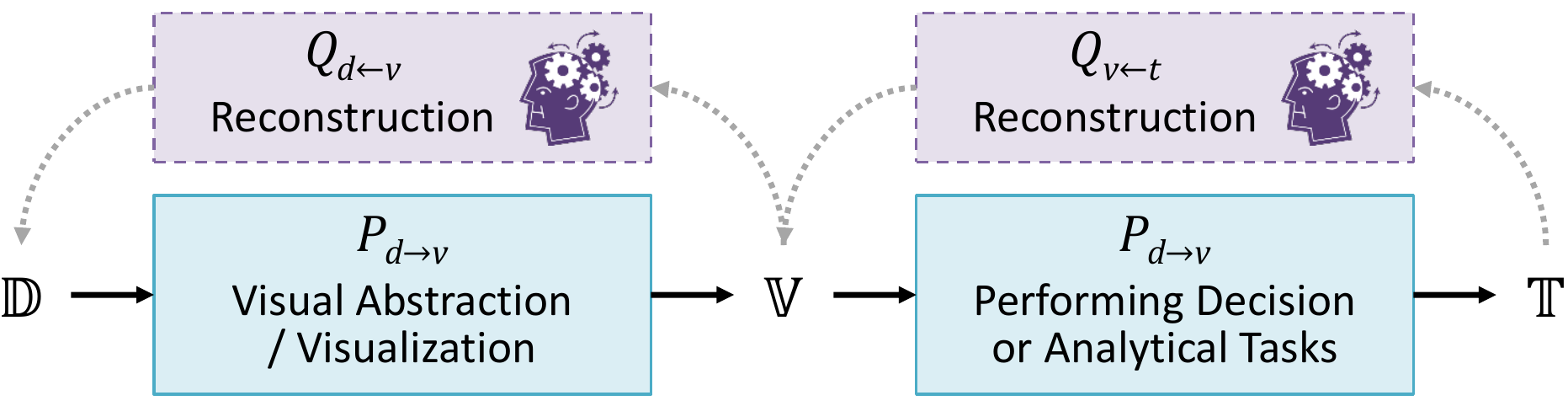}
\caption{The effectiveness of a visual abstraction process depends on the succeeding task process as well as the viewer's knowledge, biases, and cognitive capability.}
\label{fig:TwoProcesses}
\end{figure}

Consider that a viewer's interpretation is a process $Q_{d \leftarrow v} = P^{-1}_{d \rightarrow v}$ that attempts to reconstruct a dataset from a given visualization image. This process can be written as $Q_{d \leftarrow v}: \mathbb{V} \longrightarrow \mathbb{D}'$. We use $\mathbb{D}'$ to denote an alphabet that has the same set of letters as $\mathbb{D}$ but a different probability mass function from that of $\mathbb{D}$.
For example, given a bar that is 499 pixels tall, a viewer may interpret it as one of these values in the original $\mathbb{D}$, $\{498.00, 498.01, ..., 499.99, 500.00\}$. Imagine that the interpretation is biased towards 500.00 due to the corresponding mark on the vertical axis. The probability $q(500)$ would be undesirably higher than the original probability $p(500)$.

In information theory, such errors in the interpretation can be collectively measured by the Kullback-Leibler divergence, which is defined as:
\[
    \mathcal{D}_{KL}(\mathbb{D}' \| \mathbb{D}) = \sum_{d \in \mathbb{D}} q(d) \log_2 \frac{q(d)}{p(d)}
\]
where $p()$ and $q()$ are the probability mass functions of $\mathbb{D}$ and $\mathbb{D}'$ respectively, and $q(d)/p(d)$ is a discrete representation of the Radon-Nikodym derivative of $q$ with respect to the original $p$.

In the context of visual abstraction, this measurement offers a counterbalance to the measurement alphabet compression. It is referred to as \emph{Potential Distortion}~\cite{Chen:2016:TVCG:78}. While it is desirable to have the results of visual mapping $P_{d \rightarrow v}$ as abstract as possible, i.e., for $P_{d \rightarrow v}$ to have a high amount of alphabet compression, it is also necessary to keep the inaccuracy of the interpretation function $Q_{d \leftarrow v}$ as low as possible, i.e., for $Q_{d \leftarrow v}$ to have a low amount of potential distortion.

Since $Q_{d \leftarrow v}$ is a human-centric process, $Q_{d \leftarrow v}$ may feature inaccuracy due to perceptual errors and cognitive biases. However, $Q_{d \leftarrow v}$ can also make use of human knowledge that is not encoded in the data to help more accurate reconstruction. For example, imagine that a viewer is asked to guess what would be the original colors on the patch of white pixels between two black lines in the pen-and-ink visualization image in \autoref{fig:ABS-examples}(b). A na{\"i}ve guess would be either white (as what is seen) or an arbitrary selection from various grey colors.
Most viewers, especially those familiar with the depicted object or volume visualization methods, can do much better than the na{\"i}ve guess. Hence the process of ``knowledge-assisted guessing''---a heuristic process---has a lower amount of potential distortion than the na{\"i}ve guessing.
In general, it is this human knowledge that enables visual abstraction to be deployed effectively in many situations, such as those illustrated in \autoref{fig:ABS-examples}.
Whether users have the adequate ability to interpret the results of visual abstraction is thus one of the key criteria for judging if a visual abstraction process is appropriate or its results are meaningful, which reflecting the two factors (i) and (ii) described above. 

Nevertheless, since $P_{d \rightarrow v}$ is usually a many-to-one mapping, and $Q_{d \leftarrow v}$ is usually a one-to-many mapping, one may wonder why we should go through such ``unnecessary fuss'' to apply the process $P_{d \rightarrow v}$ first to $\mathbb{D}$ and another process $Q_{d \leftarrow v}$ to reconstruct $\mathbb{D}'$. One important rationale is about the \emph{task} succeeding $P_{d \rightarrow v}$ and $Q_{d \leftarrow v}$. The judgment about whether a visual abstraction process is appropriate or its results are meaningful thus depends on another process $P_{v \rightarrow t}$.
As illustrated in \autoref{fig:TwoProcesses}, $P_{v \rightarrow t}$ takes $\mathbb{V}$ as the input, and generates another output alphabet $\mathbb{T}$ that may consist of a collection of letters, e.g., different options of a decision, different levels of an assessment, different categories of a situation, etc.

The process of visual abstraction $P_{d \rightarrow v}$ and the reconstructive interpretation $Q_{d \leftarrow v}$ can collectively affect the task process $P_{v \rightarrow t}$, especially its \emph{Cost} $\mathbf{Ct}(P_{v \rightarrow t}, Q_{v \leftarrow t})$.
Similar to $Q_{d \leftarrow v}$, here $Q_{v \leftarrow t}$ is an interpretation process for reconstructing $\mathbb{V}$ from $\mathbb{T}$.
For a univariate value (e.g., 499.38), there is little merit to visualize it using a bar chart.
The difference of the cost for reading the number and that of viewing a bar is negligible for most tasks. The potential distortion caused by visual abstraction can only affect the process $P_{v \rightarrow t}$ negatively.
However, if the number of variables increases, e.g., 10 variables, the cognitive load for viewing and comparing 10 numbers is likely to be higher than viewing and comparing 10 bars using a bar chart. It is not difficult to imagine the merits of visualization when the number of variables increases. For the volume datasets featured in \autoref{fig:ABS-examples}(b,h), the number of variables in a dataset is typically at the scale of $256 \times 256 \times 256$ or more. It is inconceivable to perform a decision task by reading the numerical values of such a volume dataset. Hence, visual abstraction can be used to transform a volume dataset with a huge number of variables to visualization images as shown in (b) and (h), which reduces the cost $\mathbf{Ct}(P_{v \rightarrow t}, Q_{v \leftarrow t})$ significantly.

The above information-theoretic discourse on visual abstraction is based on the cost-benefit metric for data intelligence proposed by Chen and Golan~\cite{Chen:2016:TVCG:78}. For any data intelligence process $P_i$ with an input alphabet $\mathbb{Z}_i$ and an output alphabet $\mathbb{Z}_{i+1}$, its cost-benefit ratio is defined as:
\begin{align*}
    \frac{\textit{Benefit}}{\textit{Cost}}
    =& \frac{\textit{Alphabet Compression}-\textit{Potential Distortion}}{\textit{Cost}}\\
    =& \frac{\mathbf{AC}(P_i) - \mathbf{PD}(Q_i)}{\mathbf{Ct}(P_i, Q_i)}
    = \frac{\mathcal{H}(\mathbb{Z}_i)-\mathcal{H}(\mathbb{Z}_{i+1}) - \mathcal{D}_{KL}(\mathbb{Z}'_{i} \| \mathbb{Z}_{i})}{\mathbf{Ct}(P_i, Q_i)}
\end{align*}

When this metric is applied to the two processes in \autoref{fig:TwoProcesses}, we have the combined cost-benefit ratio as:
\begin{align} \label{eq:DVT}
    \frac{\textit{Benefit}}{\textit{Cost}} & \bigl( d \rightarrow v \rightarrow t \bigr)
    = \frac{\mathbf{AC}(P_{d \rightarrow v}) - \mathbf{PD}(Q_{d \leftarrow v})
    + \mathbf{AC}(P_{v \rightarrow t}) - \mathbf{PD}(Q_{v \leftarrow t})}%
    {\mathbf{Ct}(P_{d \rightarrow v}, Q_{d \leftarrow v})
    + \mathbf{Ct}(P_{v \rightarrow t}, Q_{v \leftarrow t})} \\
    =& \frac{\mathcal{H}(\mathbb{D})-\mathcal{H}(\mathbb{V})
        - \mathcal{D}_{KL}(\mathbb{D}' \| \mathbb{D})
        + \mathcal{H}(\mathbb{V})-\mathcal{H}(\mathbb{T})
        - \mathcal{D}_{KL}(\mathbb{V}' \| \mathbb{V})}%
    {\mathbf{Ct}(P_{d \rightarrow v}, Q_{d \leftarrow v}) + \mathbf{Ct}(P_{v \rightarrow t}, Q_{v \leftarrow t})} \nonumber\\
        =& \frac{\mathcal{H}(\mathbb{D}) - \mathcal{H}(\mathbb{T})
        - \mathcal{D}_{KL}(\mathbb{D}' \| \mathbb{D})
        - \mathcal{D}_{KL}(\mathbb{V}' \| \mathbb{V})}%
    {\mathbf{Ct}(P_{d \rightarrow v}, Q_{d \leftarrow v}) + \mathbf{Ct}(P_{v \rightarrow t}, Q_{v \leftarrow t})} \nonumber
\end{align}

\noindent In comparison, if one has to perform the task by reading the data without visualization, the cost-beneficial ratio would be:
\begin{align} \label{eq:DT}
    \frac{\textit{Benefit}}{\textit{Cost}} \bigl( d \rightarrow t \bigr)
    =& \frac{\mathbf{AC}(P_{d \rightarrow t}) - \mathbf{PD}(Q_{d \leftarrow t})}%
    {\mathbf{Ct}(P_{d \rightarrow t}, Q_{d \leftarrow t})}\\
    =& \frac{\mathcal{H}(\mathbb{D}) - \mathcal{H}(\mathbb{T})
        - \mathcal{D}_{KL}(\mathbb{D}'' \| \mathbb{D})}%
    {\mathbf{Ct}(P_{d \rightarrow t}, Q_{d \leftarrow t})} \nonumber
\end{align}

\noindent Note that the term $\mathcal{D}_{KL}(\mathbb{D}' \| \mathbb{D})$ in Eq.\,(\ref{eq:DVT}) and the term $\mathcal{D}_{KL}(\mathbb{D}'' \| \mathbb{D})$ in Eq.\,(\ref{eq:DT}) are of different quantities as they relate to $Q_{d \leftarrow v}$ and $Q_{d \leftarrow t}$ respectively. 

When the dataset $d$ is large and complex, we can see that the cost $\mathbf{Ct}(P_{d \rightarrow t}, Q_{d \leftarrow t})$ in Eq.\,(\ref{eq:DT}) would be much higher than the combined costs in Eq.\,(\ref{eq:DVT}) in terms of time and cognitive load in performing the task. In other words, we have:
\[
\mathbf{Ct}(P_{d \rightarrow t}, Q_{d \leftarrow t}) > \mathbf{Ct}(P_{d \rightarrow v}, Q_{d \leftarrow v}) + \mathbf{Ct}(P_{v \rightarrow t}, Q_{v \leftarrow t})
\]
Although reading data might appear to be more accurate, the reconstruction process $Q_{d \leftarrow t}$ from the task alphabet $\mathbb{T}$ (e.g., the patient has a tumor or not) to the data alphabet $\mathbb{D}$ (e.g., a volume dataset) is much more error-prone than the reconstruction process via visualization. In other words, we have:
\[
\mathbf{PD}(Q_{d \leftarrow t}) > \mathbf{PD}(Q_{d \leftarrow v}) + \mathbf{PD}(Q_{v \leftarrow t})
\]
With $\mathbf{AC}(P_{d \rightarrow v}) + \mathbf{AC}(P_{v \rightarrow t}) = \mathbf{AC}(P_{d \rightarrow t})$, it is not difficult to conclude:
\begin{equation*}
\frac{\textit{Benefit}}{\textit{Cost}} \bigl( d \rightarrow v \rightarrow t \bigr)
> \frac{\textit{Benefit}}{\textit{Cost}} \bigl( d \rightarrow t \bigr)
\end{equation*}

Under Condition A, we can thus mathematically reason that, for any slightly large or complex dataset, the process from data alphabet $\mathbb{D}$ to task alphabet $\mathbb{T}$ with visual abstraction is usually more cost-beneficial than the process without.

For some very simple datasets, such as a univariate value, visual abstraction may not have an information-theoretic merit. However, this is not to say that it could not have cognitive merit in disseminative visualization. More likely, the results of visual abstraction could attract more attention from the viewers who unconsciously devote more cognitive load to the task. Although the viewers' cost-beneficial ratio increases, the presenter of the disseminative visualization benefits from the contribution of extra cognitive load from the viewers. In many ways, this is similar to scenarios of disseminative visualization, where the amount of visual abstraction is purposely reduced in order to attract viewers' attention, and hence their cognitive load.
Such scenarios may include, for instance, showing an animated chart, whilst a static chart could adequately convey the information, or showing visualization in theatre-based virtual environments~\cite{Chen:2019:TVCG}.

Similarly, we can also use the cost-benefit metric to analyze the scenarios under condition B by comparing the cost-benefit ratio of a more photo-realistic technique with a less photo-realistic technique. Similar to Condition A, the potential distortion is affected by viewers' knowledge as well as their biases. The cost is affected by the viewer's task as well as cognitive capability.   

Furthermore, this metric can be applied to human-centric processes (e.g., visualization and interaction) as well as machine-centric processes (e.g., statistics and algorithms).
In general, statistical abstraction and algorithmic abstraction usually result in more alphabet compression as well as more potential distortion but less cost than visual abstraction. 
In designing a visual analytics workflow, the metric can be used to compare the cost-benefit of a human-centric process with that of a machine-centric process by
analyzing the trade-off among alphabet compression, potential distortion, and cost.
The metric can also be used to guide a visualization designer in choosing different forms of visual abstraction, e.g., in reasoning about the trade-off among the amount of abstraction, the potential perceptual errors, and the cost of task performance.

In summary, as defined at the beginning of this article, meaningful visual abstraction depends on some points of view and some tasks. From the perspective of information theory, the \textbf{points of view} may be in either or both of the following forms:
\begin{itemize}
    \item The factors that influence the \emph{alphabet compression} and \emph{cost} of the process $P_{d \rightarrow v}$ for transforming data to visualization. These factors may include the designers' wish to keep or highlight some information while removing or deemphasizing other information, their understanding of the task requirements, their appreciation of the resources available for visualization, and their awareness of the viewers' knowledge of visual representations and skills of visual analysis.
    \item The factors that influence the \emph{potential distortion} and the \emph{cost} of the process $Q_{d \leftarrow v}$ for reconstructing data from visualization. These factors may include the viewers' knowledge related to the data being depicted and the visual representations used, their understanding about the information required for performing their tasks, and their cognitive load and time constraint in executing the process $Q_{d \leftarrow v}$.
\end{itemize}
Meanwhile, \textbf{tasks} can be defined as processes that succeed the processes $P_{d \rightarrow v}$ and $Q_{d \leftarrow v}$. As long as the tasks fall broadly in the category of data intelligence tasks, the cost-benefit metric proposed by Chen and Golan~\cite{Chen:2016:TVCG:78} can also be applied to these succeeding processes.
Therefore, from the information-theoretic perspective, the most meaningful visual abstraction, or the most effective visualization in general, is the process with the optimal cost-benefit measure.  

\section{Summary}

In this article we thus formally defined the concepts of abstraction and visual abstraction as they relate to the field of visualization and based on existing notions of the terms in related fields such as the arts and in philosophy. We argued that any visual representation is the result of multiple abstraction steps from reality, and we called the step from data representation to visual representation visual abstraction. We also showed that as users of a visualization system we do not observe this abstraction process but instead adjust settings to transition from one visual representation to another---each being an independent result of the abstraction process from source thing to sign thing. Yet as designers of visualization systems we can provide guided interaction such that several results of meaningful abstractions can be assembled into sequences that we call abstraction axes to better illustrate how different aspects of reality relate to each other, and several of these abstraction axes can be assembled into abstraction spaces to illustrate the interrelation of several independent aspects. So while we argue that any visual representation is the result of an abstraction process, it is still important to discuss abstraction and visual abstraction as it teaches us about visualization as a process in general.

\section*{Acknowledgments}
The authors would like to thank Jos Roerdink, Helwig Hauser, Stefan Bruckner, Hans-Christian Hege, and Torsten M\"oller for fruitful discussion that helped shaping the article. Thanks to Peter Mindek for illustrating the abstraction space in \autoref{fig:space}. This work was funded through the ILLUSTRARE grant by
both the Austrian Science Fund (FWF; I 2953-N31) and the French National Research Agency (ANR; ANR-16-CE91-0011-01).



\bibliographystyle{unsrt}  
\bibliography{paper}  

\end{document}